\documentclass[12pt,reqno]{amsart}

\usepackage{times}
\usepackage{graphicx}
\graphicspath{ {./images/} }

\usepackage{geometry}

\usepackage{amsmath,amsthm,amsfonts,amscd,amssymb}
\usepackage{pst-node}
\usepackage{pst-plot}
\usepackage[all]{xy}
\usepackage{stackengine}
\usepackage{multirow}

\usepackage{tikz, braids}
\usetikzlibrary{decorations.pathreplacing, arrows}

\usepackage[most]{tcolorbox}

\newtcolorbox{mainresultbox}{
    colback=white,
    colframe=black,
    boxrule=0.8pt,
    arc=1mm,
    left=6pt,
    right=6pt,
    top=6pt,
    bottom=6pt,
    before skip=10pt,
    after skip=10pt
}

\newtheorem{theorem}{Theorem}[section]
\newtheorem{corollary}[theorem]{Corollary}

\newtheorem{lemma}[theorem]{Lemma}
\newtheorem{remark}[theorem]{Remark}
\newtheorem{proposition}[theorem]{Proposition}

\newtheorem*{maintheorem}{Main Theorem}

\numberwithin{theorem}{section}
\numberwithin{equation}{section}

\newcommand{\Aa}{\mathcal{A}}

\newcommand{\la}{\langle}
\newcommand{\ra}{\rangle}

\newcommand{\re}{\mathbb{R}}

\newcommand{\om}{\omega}

\begin{document}

\makeatletter

\def\@settitle{%
  \begin{center}%
    \baselineskip14\p@\relax
    {\LARGE\bfseries \@title\par}%
  \end{center}%
}

\makeatother

\title[An optimal quantum strategy for the $2$-hat game]
{An optimal quantum strategy for the $2$-hat game}

\author[Georgios Baziotis]{Georgios Baziotis}

\address{Georgios Baziotis,
Research Institute of Mathematics and NextQuantum, Seoul National University, Gwanak-Ro 1, Gwanak-Gu, Seoul 08826, Republic of Korea}

\email{baziotis@snu.ac.kr}

\author[Sang-Jun Park]{Sang-Jun Park}

\address{Sang-Jun Park,
School of Mathematics and Statistics, Wuhan University, 430070, Wuhan, Hubei, China}

\email{sjpark@whu.edu.cn}

\author[Sang-Gyun Youn]{Sang-Gyun Youn}

\address{Sang-Gyun Youn,
Department of Mathematics Education and NextQuantum, Seoul National University, Gwanak-Ro 1, Gwanak-Gu, Seoul 08826, Republic of Korea}

\email{s.youn@snu.ac.kr}

\begin{abstract}
We study the two-player hat game as a two-input, three-output non-local game. We calculate the quantum commuting value and show that it coincides with the quantum one. In particular, we show that 
\[\omega_{\rm q}(H_2)=\omega_{\mathrm{qc}}(H_2)=\frac{15+\sqrt{33}}{36}\approx 0.576238.\]
We achieve the optimal quantum advantage using a shared non-maximally entangled qutrit state. To establish optimality, we exploit the game's symmetry under the dihedral group $D_8$ and apply its representation theory to substantially reduce the complexity of the underlying matrix problem. We also determine the no-signalling value $\omega_{\mathrm{ns}}(H_2)=2/3$.
\end{abstract}
\maketitle
\section{Introduction}
In recent years, non-local games have emerged as a key framework in understanding the peculiar phenomena of entanglement and non-locality. Originally introduced in the context of quantum information theory \cite{ClHoToWa04}, they are nowadays used to study Bell inequalities, complexity and structural questions in operator algebras \cite{JiNaViWrYu20,JuNaPaPGScWe11,Fritz12,Sl20}. Moreover, particular games have revealed entanglement-assisted structures in both classical and non-commutative objects \cite{LuMaPaRoScSeToWi20,PaTo15,AtMaRoSaSeVa19,BrChEiHaPaSuWa20}.
\par Throughout this paper we are primarily concerned with cooperative $n$-hat games and their quantum values. The classical $n$-hat game $H_n$ is a cooperative guessing problem involving $n$ players, each of whom wears a hat whose color is chosen independently and uniformly from two possibilities. Every player can see the hats of the other $n-1$ players but not their own. Before the game begins, the players may agree upon a common strategy; once the hats have been assigned, no further communication is allowed. Each player must either guess the color of their own hat or pass. The team wins if at least one player makes a guess and every guess made is correct. Since Ebert's doctoral dissertation \cite{Eb98}, this problem has attracted considerable attention and has revealed close connections with coding theory and theoretical computer science. Deterministic strategies can be described in terms of covering codes of radius one, with the losing configurations forming the code \cite{LeSe02}. In particular, when $n=2^m-1$, a perfect Hamming code gives a strategy attaining the general upper bound $n/(n+1)$; for $n=3$, this construction reduces to the familiar strategy associated with the repetition code $\{000,111\}$. The problem has also appeared in the study of autoreducibility of random sequences \cite{EbVo00,EbMeVo03}, and numerous variations involving different winning conditions, more than two hat colors, or restricted visibility among the players have been investigated \cite{Kr10,PaSt10,Ta14,BuHaKlLe08,Fe10,HoKr12}. Related graph-based guessing games have further been connected with graph entropy, network coding, index coding, locally recoverable distributed storage, and finite dynamical systems \cite{GaRi11,ArKi15,Ga18}.

In this paper, we focus on the two-player case, which we view as a non-local game and continue to denote by $H_2$. Alice and Bob wear hats whose colors are represented by independent and uniformly distributed bits $x,y\in\{0,1\}$, respectively. Alice observes $y$, Bob observes $x$, and each player produces one of three outputs: guess $0$, guess $1$, or pass. They win if Alice guesses $x$ and Bob passes, if Alice passes and Bob guesses $y$, or if they both guess correctly. They lose if they both
pass or if either player makes an incorrect guess.  
The classical two-player hat game has a particularly simple optimal strategy. 
Classically, the optimal winning probability is
$\omega_{\rm loc}(H_2)=\frac12$,
attained, for example, by the strategy in which one player always guesses and the other always passes.

The situation changes when the players are allowed to share entanglement. After observing the other player's hat color, Alice and Bob may perform local three-outcome measurements with outputs $0$, $1$, and $P$, where $P$ denotes pass. In this formulation, the two-player hat game is a two-input, three-output non-local game. Quantum strategies for $n$-hat games were previously considered by Zhao, Chen, Xu, and Yang \cite{ZhChXuYa22}, who proposed constructions based on quantum stabilizer codes. For $n=2,3,4$, their strategies attain winning probabilities $1/2$, $3/4$, and $3/4$, respectively. These coincide with the corresponding optimal classical winning probabilities \cite{LeSe02,GuKaRaTu06} and hence do not yield a quantum advantage in these cases.

A quantum advantage for the two-player hat game follows from the recent work of Schmidt, Storgaard, Walter, and Zhao on Feige's game \cite{ScStWaZh25}. Feige's game has the same input and output sets, but its winning condition is stricter: exactly one player must guess correctly while the other passes. In particular, the players lose when
they both guess correctly. The quantum strategy constructed in \cite{ScStWaZh25} wins Feige's game with probability $9/16$. Applied without modification to the two-player hat game, the same strategy also wins on the additional event in which both players guess correctly, and hence achieves
\begin{align}
\frac{9}{32}+\frac{9}{32}+\frac{1}{192}=\frac{109}{192} \approx 0.567708.
\end{align}
This already establishes a genuine quantum advantage in the two-player hat game.

In this paper, we show that $109/192$ is not optimal among quantum strategies. We write $\omega_{\rm q}$ for the supremum of the winning probability using local measurements on a shared finite-dimensional quantum state. The quantum commuting value $\omega_{\mathrm{qc}}$ allows measurement operators on a common Hilbert space of arbitrary dimension, with operators belonging to different players required to commute. Our main result is the following.
\begin{mainresultbox}
\begin{maintheorem}
For the two-player hat game $H_2$,
\begin{equation}
\omega_{\rm q}(H_2)=\omega_{\mathrm{qc}}(H_2)
=\frac{15+\sqrt{33}}{36}\approx 0.576238.
\end{equation}
\end{maintheorem}
\end{mainresultbox}

The definitions of non-local games and their strategies are recalled in Section~\ref{sec-nlgames}. In Section~\ref{sec-lower}, we construct an explicit strategy attaining the value in the Main Theorem, using a non-maximally entangled qutrit state and rank-one projective measurements.

In Section~\ref{sec-upper}, we prove the matching upper bound for $\omega_{\mathrm{qc}}(H_2)$. The key idea is to identify the dihedral group $D_8$ as a symmetry group of $H_2$. We then apply its representation theory to substantially reduce the computational complexity of the underlying matrix problem. This reduction allows us to construct an exact certificate for the upper bound. Appendix \ref{app:representations} gives the detailed representation-theoretic calculations. Appendix \ref{app:certificate} presents the explicit construction of the certificate and the calculations needed to verify it rigorously and independently.

A natural question is whether quantum advantage can occur more generally in the $n$-player hat game. To study this question, we consider the broader class of no-signalling strategies, which contains all quantum commuting strategies. We denote the supremum of the winning probability over this class by $\omega_{\mathrm{ns}}$. In Section \ref{sec-general-upper-bound}, we determine $\omega_{\mathrm{ns}}(H_2)=2/3$ and show that the classical upper bound extends to all no-signalling strategies:
\[\omega_{\mathrm{ns}}(H_n)\leq\frac{n}{n+1}, \qquad n\geq1.\]
Consequently, for $n=2^m-1$ with $m\geq1$, the known classical strategies attaining this bound yield
\[\omega_{\mathrm{loc}}(H_n)=\omega_{\rm q}(H_n)=\omega_{\mathrm{qc}}(H_n)=\omega_{\mathrm{ns}}(H_n)=\frac{n}{n+1}.\]
Thus, for these values of $n$, even no-signalling strategies offer no advantage over classical strategies.

\subsection*{AI tools disclosure.}
In Section 3, we first hypothesized an asymmetric structure for the quantum state and specific forms for the measurements. We parameterized both to design a numerical algorithm and identified candidate optimal parameter values through numerical experiments assisted by OpenAI's GPT-6 Astra. In Section~4, we applied representation theory to design numerical experiments with substantially reduced computational complexity and used the same tool to obtain numerical candidates for the real symmetric positive semidefinite blocks presented in Proposition \ref{prop:exact-certificate}. We organized and simplified the resulting constructions and verified them by exact calculations. We take full responsibility for the content of this paper.

\section{Non-local games and their strategies}\label{sec-nlgames}
A non-local game \cite{ClHoToWa04} is a tuple $\mathcal{G}=(X,Y,A,B,\pi,V)$, where $X,Y,A,B$ are finite sets, $\pi$ is a probability measure on $X\times Y$ and $V:X\times Y\times A\times B\rightarrow\{0,1\}$ is the rule function. The game proceeds as follows. In each round, the verifier selects a pair of questions $(x,y)\in X\times Y$ according to $\pi$ and sends $x$ and $y$ to Alice and Bob, respectively. Upon receiving their questions, the players respond with answers $a\in A$ and $b\in B$. The verifier uses the rule function $V$ to determine the outcome: the players win if $V(x,y,a,b)=1$ and lose if $V(x,y,a,b)=0$. The players may agree on a joint strategy before the game begins, but they are not allowed to communicate during the game. {Each strategy between Alice and Bob can then be modeled as a family of conditional probabilities $p=\big(p(\cdot,\cdot|x,y)\big)_{(x,y)\in X\times Y}$ whose marginals satisfy the no-signalling conditions: for all $x,x'\in X$ and $y,y'\in Y$,
\begin{align*}
    \sum_{a\in A}p(a,b|x,y)&=\sum_{a\in A}p(a,b|x',y) \;\forall\,b\in B, \text{ and}\\
    \sum_{b\in B}p(a,b|x,y)&=\sum_{b\in B}p(a,b|x,y') \;\forall\,a\in A.
\end{align*}
We call $p$ satisfying the above conditions a \emph{no-signalling} strategy for $\mathcal{G}$ and define the \emph{average winning probability} under $p$ as
\begin{equation} \label{eq:winningprob}
    \omega(\mathcal{G},p):=\sum_{(x,y,a,b)\in X\times Y\times A\times B}\pi(x,y)V(x,y,a,b)p(a,b|x,y).
\end{equation}

} 

Strategies corresponding to classical mechanics are called \emph{local} and are convex combinations of \emph{deterministic} strategies. A deterministic strategy is specified by a pair of functions $f:X\rightarrow A$ and $g:Y\rightarrow B$ via $p(a,b|x,y)=\mathbf{1}_{a,f(x)}\mathbf{1}_{b,g(y)}$ for all $(x,y,a,b)\in X\times Y\times A\times B.$ The sets of all no-signalling and local correlations are denoted by $\mathcal{C}_{\rm ns}$ and $\mathcal{C}_{\rm loc}$, respectively.

We denote by $\mathcal{H}$ a Hilbert space of arbitrary dimension (unless stated otherwise), and write $\mathcal{B}(\mathcal{H})$ for the $C^*$-algebra of all bounded linear operators on $\mathcal H$. A $\mathcal{B}(\mathcal{H})$-valued \emph{positive operator-valued measure (POVM)} on $A$ is a collection $(M_a)_{a\in A}$ of positive operators such that $\sum_{a\in A}M_a=I_\mathcal{H}$. Moreover, if $M_aM_{a'}=\delta_{a,a'}M_a$, then $(M_a)_{a\in A}$ is called a \emph{projection-valued measure (PVM)}. A strategy $p$ is called \emph{quantum} if there exist finite-dimensional Hilbert spaces $\mathcal{H}$ and $\mathcal{K}$, a unit vector $|\Omega\rangle\in \mathcal{H}\otimes \mathcal{K}$ and PVMs $(M^x_a)_{a\in A}\subseteq\mathcal{B}(\mathcal{H})$ and $(N^y_b)_{b\in B}\subseteq \mathcal{B}(\mathcal{K})$ for all $(x,y)\in X\times Y$, such that $$p(a,b|x,y)=\langle\Omega| M^x_a\otimes N^y_b|\Omega\rangle, \ \ (x,y,a,b)\in X\times Y\times A\times B.$$
We denote by $\mathcal{C}_{\rm q}$ the set of all quantum correlations. %
Given two $\mathcal{B}(\mathcal{H})$-valued POVMs, $\mathcal{M}=(M_a)_{a\in A}$ on $A$ and $\mathcal{N}=(N_b)_{b\in B}$ on $B$, we say that $\mathcal{M}$ and $\mathcal{N}$ have commuting ranges if $M_aN_b=N_bM_a$ for all $(a,b)\in A\times B$. A correlation $p$ is called \emph{quantum commuting} if there exist a Hilbert space $\mathcal{H}$ and PVMs $\mathcal{M}^x=(M^x_a)_{a\in A}\subseteq\mathcal{B}(\mathcal{H})$ and  $\mathcal{N}^y=(N^y_b)_{b\in B}\subseteq\mathcal{B}(\mathcal{H})$ for all $(x,y)\in X\times Y$ with commuting ranges and a unit vector $|\Omega\rangle\in \mathcal{H}$, such that 
$$p(a,b|x,y)=\langle\Omega| M^x_a N^y_b|\Omega\rangle, \ \ (x,y,a,b)\in X\times Y\times A\times B. $$
The set of all quantum commuting correlations is denoted by $\mathcal{C}_{\rm qc}$. Moreover, we note that by a dilation argument \cite[Remark 11]{JuNaPaPGScWe11}, the use of PVMs in the definitions of both quantum and quantum commuting correlations is equivalent to the POVM formulation. The following inclusions hold:
\begin{align}\label{chain_corr}
    \mathcal{C}_{\rm loc}\subseteq\mathcal{C}_{\rm q}\subseteq \mathcal{C}_{\rm qc}\subseteq\mathcal{C}_{\rm ns},
\end{align}
\par One way to distinguish the correlation classes in \eqref{chain_corr} is to calculate the value of a game, namely the optimal winning probability for the game $\mathcal{G}$. For ${\rm t}\in \{\rm loc, q, qc, ns\}$, the optimal winning probability of the players using classical (resp. quantum, quantum commuting, no-signalling) resources is called the \emph{$t$-value of the game} and is given by the quantity 
\begin{align*}
    \omega_{\rm t}(\mathcal{G}) := \sup_{p\in \mathcal{C}_{\rm t}} \om(\mathcal{G},p)=\sup_{p\in \mathcal{C}_{\rm t}}\sum_{x,y,a,b}\pi(x,y)V(x,y,a,b)p(a,b|x,y).
\end{align*}
In this context, the inclusions in \eqref{chain_corr} give the corresponding inequalities between game values: $$\omega_{\rm loc}(\mathcal{G})\leq\omega_{\rm q}(\mathcal{G})\leq \omega_{\rm qc}(\mathcal{G})\leq \omega_{\rm ns}(\mathcal{G}).$$

For $n$ players, local correlations are convex combinations of products of deterministic response functions. Quantum correlations are defined using one tensor factor for each player, while quantum commuting correlations are defined using measurement operators that commute whenever they belong to distinct players. We use the same notation for the corresponding $n$-player game values.

\section{A lower bound for $\omega_{\rm q}(H_2)$}\label{sec-lower}

In this section, we present the optimal quantum strategy and calculate its value. As in the introduction, $x$ and $y$ denote Alice's and Bob's hat colors, so Alice receives $y$ and Bob receives $x$; accordingly, we write the correlation as $p(a,b|y,x)$. Let $\mathcal{H}=\mathcal{K}=\mathbb{C}^3$, and suppose that Alice and Bob share the bipartite pure state $|\Omega\rangle\in\mathbb C^3\otimes\mathbb C^3$ given by
\begin{align*}
|\Omega\ra=
\alpha|00\ra+\alpha|11\ra+\beta|22\ra=
\alpha\bigl(|00\ra+|11\ra+t|22\ra\bigr)
\end{align*}
with the coefficients
\begin{align*}
\alpha&=\sqrt{\frac{11+\sqrt{33}}{44}},
&\beta&=\sqrt{\frac{11-\sqrt{33}}{22}},
&t&=\frac{\beta}{\alpha}
=\sqrt{\frac{7-\sqrt{33}}{2}}.
\end{align*}
The normalization follows from $\alpha^2(2+t^2)=1$. To describe Alice's measurement for input $y$ and Bob's measurement
for input $x$, define, for $\varphi\in[0,\pi]$ and $\theta\in\mathbb{R}$,
\begin{align*}
u(\varphi,\theta)=\begin{bmatrix}
\sin(\varphi)\cos(\theta)\\
\sin(\varphi)\sin(\theta)\\
\cos(\varphi)
\end{bmatrix},
\end{align*}
and let $\gamma= \arccos\!\left(\frac{1}{\sqrt{3}}\right)\in \left(\frac{\pi}{4},\frac{\pi}{3}\right)$.
For any $y\in \{0,1\}$ and $a\in\{0,1,P\}$, we define Alice's rank-one projections by
$M_a^y=|u_a^y\ra\la u_a^y|$, where
{
\begin{align*}
u_0^0:= u\left(\gamma,\frac{11\pi}{6}\right)= \begin{bmatrix} \frac{1}{\sqrt{2}}\\ -\frac{1}{\sqrt{6}}\\ \frac{1}{\sqrt{3}} \end{bmatrix},&\quad u_1^0:=
 u\left(\pi-\gamma,\frac{\pi}{6}\right) = \begin{bmatrix} \frac{1}{\sqrt{2}}\\ \frac{1}{\sqrt{6}}\\ -\frac{1}{\sqrt{3}} \end{bmatrix},\\[2mm]
u_P^0:=
 u\left(\gamma,\frac{\pi}{2}\right) = \begin{bmatrix} 0\\ \frac{2}{\sqrt{6}}\\ \frac{1}{\sqrt{3}} \end{bmatrix},&\quad u_0^1:=u\left(\pi-\gamma,\frac{7\pi}{6}\right) = \begin{bmatrix} -\frac{1}{\sqrt{2}}\\ -\frac{1}{\sqrt{6}}\\ -\frac{1}{\sqrt{3}} \end{bmatrix},\\[2mm]
u_1^1:=u\left(\pi-\gamma,\frac{11\pi}{6}\right) = \begin{bmatrix} \frac{1}{\sqrt{2}}\\ -\frac{1}{\sqrt{6}}\\ -\frac{1}{\sqrt{3}} \end{bmatrix},&\quad u_P^1:= u\left(\gamma,\frac{3\pi}{2}\right) = \begin{bmatrix} 0\\ -\frac{2}{\sqrt{6}}\\ \frac{1}{\sqrt{3}} \end{bmatrix}.
\end{align*}
}
Similarly, for $x\in\{0,1\}$ and $b\in\{0,1,P\}$, we define Bob's rank-one projections by $N_b^x=|v_b^x\ra\la v_b^x|$, where
{\small
\begin{align*}
v_0^0:= u\left(\pi-\gamma,\frac{5\pi}{3}\right) = \begin{bmatrix} \frac{1}{\sqrt{6}}\\ -\frac{1}{\sqrt{2}}\\ -\frac{1}{\sqrt{3}} \end{bmatrix},&\quad v_1^0:=
 u\left(\pi-\gamma,\frac{\pi}{3}\right) = \begin{bmatrix} \frac{1}{\sqrt{6}}\\ \frac{1}{\sqrt{2}}\\ -\frac{1}{\sqrt{3}} \end{bmatrix},\\[2mm]
v_P^0:=
  u\left(\pi-\gamma,\pi\right) = \begin{bmatrix} -\frac{2}{\sqrt{6}}\\ 0\\ -\frac{1}{\sqrt{3}} \end{bmatrix},&\quad v_0^1:=u\left(\pi-\gamma,\frac{4\pi}{3}\right) = \begin{bmatrix} -\frac{1}{\sqrt{6}}\\ -\frac{1}{\sqrt{2}}\\ -\frac{1}{\sqrt{3}} \end{bmatrix},\\[2mm]
v_1^1:=u\left(\pi-\gamma,\frac{2\pi}{3}\right) = \begin{bmatrix} -\frac{1}{\sqrt{6}}\\ \frac{1}{\sqrt{2}}\\ -\frac{1}{\sqrt{3}} \end{bmatrix},&\quad v_P^1:= u\left(\gamma,\pi\right) = \begin{bmatrix} -\frac{2}{\sqrt{6}}\\ 0\\ \frac{1}{\sqrt{3}} \end{bmatrix},
\end{align*}
}

For each input, the three measurement vectors form an orthonormal basis of $\mathbb C^3$. We denote the corresponding PVMs by
\[
\mathcal M^y=(M_a^y)_{a\in\{0,1,P\}},\qquad
\mathcal N^x=(N_b^x)_{b\in\{0,1,P\}}.
\]

Recall that, given a strategy $p\in \mathcal{C}_{\rm ns}$, the average winning probability \eqref{eq:winningprob} under $p$ is $$\omega(H_2,p)=\sum_{x,y,a,b}\pi(y,x)V(y,x,a,b)p(a,b|y,x).$$
For the strategy defined above, straightforward calculations show that 
\begin{align*}
\omega(H_2,p)=\sum_{x,y}\frac{1}{4}\Big(\left|\la\Omega|u_x^y\otimes v_P^x\ra\right|^2+\left|\la\Omega|u_P^y\otimes v_y^x\ra\right|^2+\left|\la\Omega|u_x^y\otimes v_y^x\ra\right|^2\Big).
\end{align*}
Let $D_t=\operatorname{diag}(1,1,t)$ and note that for any $\xi,\eta\in\mathbb{C}^3$, we have
$\la\Omega|\xi\otimes\eta\ra=\alpha \cdot \xi^T D_t\eta.$

\begin{proposition}\label{prop:quantum-lower-bound}
The state $|\Omega\rangle$ and the projective measurements $\mathcal M^y$ and $\mathcal N^x$ defined above give a quantum strategy for $H_2$ with winning probability $(15+\sqrt{33})/36$. In particular,
\[
\omega_{\rm q}(H_2)\geq\frac{15+\sqrt{33}}{36}.
\]
\end{proposition}
\begin{proof}
A direct calculation gives
\begin{align}
\begin{array}{c|c|c|c}
xy&\left(u_x^y\right)^TD_tv_P^x&\left(u_P^y\right)^TD_tv_y^x&\left(u_x^y\right)^TD_tv_y^x\\ 
\hline
00&-\dfrac{t+\sqrt{3}}{3}&-\dfrac{t+\sqrt{3}}{3}&\dfrac{\sqrt{3}-t}{3}\\[3mm]
01&\dfrac{t+\sqrt{3}}{3}&-\dfrac{t+\sqrt{3}}{3}&\dfrac{t-\sqrt{3}}{3}\\[3mm]
10&-\dfrac{t+\sqrt{3}}{3}&-\dfrac{t+\sqrt{3}}{3}&\dfrac{t-\sqrt{3}}{3}\\[3mm]
11&-\dfrac{t+\sqrt{3}}{3}&-\dfrac{t+\sqrt{3}}{3}&\dfrac{t-\sqrt{3}}{3}
\end{array}\label{tab:corr}    
\end{align}

By \eqref{tab:corr}, for all $xy\in\{0,1\}^2$, we obtain
$$p(x,P|y,x)=p(P,y|y,x)=\alpha^2\frac{(t+\sqrt{3})^2}{9}=\frac{55+9\sqrt{33}}{396},$$
and
$$p(x,y|y,x)=\alpha^2\frac{(t-\sqrt{3})^2}{9}=\frac{55-7\sqrt{33}}{396}.$$
Consequently,
\begin{align*}
\omega(H_2,p)=\sum_{x,y}\frac{1}{4}\left( 2\left(\frac{55+9\sqrt{33}}{396}\right)+\frac{55-7\sqrt{33}}{396}\right)=\frac{15+\sqrt{33}}{36},
\end{align*}
establishing a quantum advantage for $H_2$ and providing a lower bound for $\omega_{\rm q}(H_2)$ and $\omega_{\rm qc}(H_2).$
\end{proof}

\section{An upper bound for $\omega_{\mathrm{qc}}(H_2)$}\label{sec-upper}
In this section, we obtain an upper bound for the quantum commuting value of $H_2$ and show that it coincides with the value of the quantum strategy introduced in Section \ref{sec-lower}. Before proceeding, we recall some useful operator-algebraic interpretations of quantum commuting correlations introduced in \cite{Fritz12,JuNaPaPGScWe11} (see also \cite{LuMaPaRoScSeToWi20}).

\par For $H_2$, we set $X=Y=\{0,1\}$ and $A=B=\{0,1,P\}$. The following facts hold for arbitrary finite question and answer sets. Let $\ast_{x\in X}\mathbb{Z}_{\lvert A\rvert}$ denote the free product of $\lvert X\rvert$ copies of $\mathbb{Z}_{\lvert A\rvert}$, and write $\mathcal{A}_{X,A}=\rm{C}^*(\ast_{x\in X}\mathbb{Z}_{\lvert A\rvert})$ for the full group $C^*$-algebra of $\ast_{x\in X}\mathbb{Z}_{\lvert A\rvert}$. Given a state $s:\mathcal{A}_{X,A}\otimes_{\rm max}\mathcal{A}_{Y,B}\rightarrow \mathbb{C}$, one obtains a correlation $p_s$ via 
\begin{align}\label{eq:state_corr}
    p_s(a,b|x,y)=s(e_{x,a}\otimes f_{y,b}), \ (a,b,x,y)\in A\times B\times X\times Y,
\end{align}
where $(e_{x,a})_{(x,a)\in X\times A}$ (resp. $(f_{y,b})_{(y,b)\in Y\times B}$) denote the canonical measurement projections of $\mathcal{A}_{X,A}$ (resp. $\mathcal{A}_{Y,B}$) and $\otimes_{\rm max}$ denotes the maximal $C^*$-algebraic tensor product (see \cite{BrOz08}). Then by considering the GNS representation of $\Aa$, we deduce that $p_s$ defines a quantum commuting correlation. Conversely, given a quantum commuting correlation $p$, the universal property of $\Aa$ implies that there exists a state $s_p$ satisfying \eqref{eq:state_corr}. Thus $s\mapsto p_s$ defines an affine surjection from the state space of $\mathcal{A}_{X,A}\otimes_{\rm max}\mathcal{A}_{Y,B}$ onto $\mathcal{C}_{\rm qc}$ \cite[Proposition~3.4(b)]{Fritz12}. %
\par The $C^*$-algebra $\mathcal{A}=\mathcal{A}_{X,A}\otimes_{\rm max}\mathcal{A}_{Y,B}$
serves as an ambient space for \emph{Bell operators}. Recall that given a non-local game $\mathcal{G}=(X,Y,A,B,\pi,V)$, a Bell operator $\Phi_{\mathcal{G}}$ is an element of $\mathcal{A}$ given by $$\Phi_{\mathcal{G}}=\sum_{x,y,a,b} \pi (x,y)V(x,y,a,b)e_{x,a}\otimes f_{y,b}.$$
This formulation yields upper bounds for the quantum commuting value of $\mathcal{G}$ through sum of squares certificates. Namely, given $\lambda>0$, consider the element $\lambda I-\Phi_\mathcal{G}$ and note that for any state $s:\mathcal{A}_{X,A}\otimes_{\rm max}\mathcal{A}_{Y,B}\rightarrow \mathbb{C}$, one obtains 
\begin{align}\label{eq:bell}
s(\lambda I-\Phi_\mathcal{G})=\lambda-\sum_{x,y,a,b}\pi(x,y)V(x,y,a,b)s(e_{x,a}\otimes f_{y,b})=\lambda-\omega(\mathcal{G},p_s).
\end{align}
Identity \eqref{eq:bell} shows that $\lambda$ is an upper bound for the quantum commuting value whenever $\lambda I-\Phi_{\mathcal{G}}$ is a positive element in $\mathcal{A}_{X,A}\otimes_{\rm max}\mathcal{A}_{Y,B}$. To establish this positivity, it suffices to find elements $c_1,\dots,c_n\in \mathcal{A}_{X,A}\otimes_{\rm max}\mathcal{A}_{Y,B}$, satisfying $$\lambda I-\Phi_{\mathcal{G}}=\sum_{i=1}^nc_i^*c_i\geq0.$$
We use this approach to calculate the quantum commuting value of $H_2$.
\subsection{The upper bound}\label{sec:bell-operator}
Set $\displaystyle \lambda=\frac{15+\sqrt{33}}{36}$. We first describe the Bell operator for $H_2$ and explain how a positive semidefinite matrix certificate would imply $\lambda I-\Phi_{H_2}\geq0$ in $\mathcal{A}$. For this purpose, recall that the $C^*$-algebra $\mathcal{A}=\mathcal{A}_{X,A}\otimes_{\rm max}\mathcal{A}_{Y,B}$ is the universal $C^*$-algebra generated by projections 
\begin{align*}
\{M_0^y,M_1^y,M_P^y,N_0^x,N_1^x,N_P^x\}, \ \ x,y\in\{0,1\}    
\end{align*}
subject to the relations
\begin{itemize}
    \item[(i)] $M_a^yM_{a'}^y= \delta_{a,a'}M_a^y$ and $\sum_{a\in\{0,1,P\}}M_a^y=I$;
    \item[(ii)] $N_b^xN_{b'}^x=\delta_{b,b'}N_b^x$ and $\sum_{b\in\{0,1,P\}}N_b^x=I$;
    \item[(iii)] $[M_a^y,N_b^x]=0$
\end{itemize}
for all $x,y\in\{0,1\}$ and $a,a',b,b'\in\{0,1,P\}$. We refer to (i) and (ii) as the PVM relations and to (iii) as the commuting relations. To construct the exact certificate, we first express $\Phi_{H_2}$ without the generators corresponding to the output $P$.
\begin{proposition}
The Bell operator $\Phi_{H_2}$ for the $2$-hat game $H_2$ is given by 
\begin{align}\label{eq30}
\Phi_{H_2}=\frac14\sum_{x,y\in\{0,1\}}\Bigl(&M_x^y+N_y^x-M_x^yN_y^x-M_x^yN_{1-y}^x-M_{1-x}^yN_y^x\Bigr).
\end{align}
\end{proposition}
\begin{proof}
    First note that for the hat configuration $(x,y)$, the three winning output pairs are $(x,P)$, $ (P,y),$ and $(x,y)$. Averaging over the four hat configurations gives the Bell operator
\begin{align*}
\Phi_{H_2}=\frac14 \sum_{x,y\in\{0,1\}} \left(M_x^yN_P^x+M_P^yN_y^x+M_x^yN_y^x\right).
\end{align*}
 Using the PVM relations to eliminate the pass projections gives
 \begin{align*}
M_x^yN_P^x&=M_x^y-M_x^yN_y^x-M_x^yN_{1-y}^x,& 
M_P^yN_y^x&=N_y^x-M_x^yN_y^x-M_{1-x}^yN_y^x,
\end{align*}
and substitution yields \eqref{eq30}.
\end{proof}
Henceforth, we use the expression for $\Phi_{H_2}$ in \eqref{eq30}.

To obtain the desired upper bound, it is enough to find a column $\mathbf w\in M_{25,1}(\mathcal A)$ of operator words and a real symmetric matrix $Z\geq0$ satisfying
\begin{equation}\label{eq:certificate-criterion}
\lambda I-\Phi_{H_2}=\mathbf w^*(Z\otimes1)\mathbf w.
\end{equation}
If such a matrix $Z$ exists, then $Z=R^TR$ for some real matrix $R$, and hence
\begin{align*}
\lambda I-\Phi_{H_2}
&=\mathbf w^*(R^TR\otimes1)\mathbf w\\
&=\bigl((R\otimes1)\mathbf w\bigr)^*\bigl((R\otimes1)\mathbf w\bigr)\geq0.
\end{align*}
Thus $\Phi_{H_2}\leq\lambda I$, so every commuting-operator strategy has winning probability at most $\lambda$. The existence of such a certificate would therefore imply
\[
\omega_{\mathrm{qc}}(H_2)\leq\frac{15+\sqrt{33}}{36}.
\]
Together with Proposition~\ref{prop:quantum-lower-bound}, this would prove the Main Theorem. We specify $\mathbf w$ in Section~\ref{sec-upper1} and establish the existence of such a matrix $Z$ in Section~\ref{sec-upper2}.

\subsection{The coefficient equations} \label{sec-upper1}
To express the certificate identity as a system of coefficient equations, fix the ordered column vector of operator words
\begin{align}
\mathbf w = \Bigl(
&I,
M_0^0,M_1^0,M_0^1,M_1^1,
N_0^0,N_1^0,N_0^1,N_1^1,
M_0^0N_0^0,M_0^0N_1^0,
M_0^0N_0^1,M_0^0N_1^1,
\notag\\
&
M_1^0N_0^0,M_1^0N_1^0,
M_1^0N_0^1,M_1^0N_1^1,
M_0^1N_0^0,M_0^1N_1^0,
M_0^1N_0^1,M_0^1N_1^1,
\notag\\
&
M_1^1N_0^0,M_1^1N_1^0,
M_1^1N_0^1,M_1^1N_1^1
\Bigr)^T
\label{eq:word-vector},
\end{align}
and write ${\bf w}=(w_j)_{j=0}^{24}=\displaystyle \sum_{j=0}^{24}|j\ra\otimes w_j\in M_{25,1}(\mathcal{A})$, where the last sixteen words are ordered lexicographically in $(y,a,x,b)$. For this choice of $\mathbf w$, we seek a positive semidefinite matrix $Z\in M_{25}$ satisfying
\begin{align}\label{eq301}
\lambda I-\Phi_{H_2}=\mathbf w^*(Z\otimes1)\mathbf w = \sum_{i,j=0}^{24}z_{ij}w_i^*w_j.
\end{align} 
\begin{remark} \rm If a (complex) positive semidefinite matrix $Z$ satisfies \eqref{eq301}, then so does $\overline Z$, since the defining relations and all coefficients of $\Phi_{H_2}$ and $\mathbf w$ are real. Thus, $\frac12(Z+\overline Z)=\operatorname{Re}Z$ is a positive semidefinite solution. Hence, we may restrict attention to real symmetric matrices $Z\in M_{25}(\re)_{sym}$. 
\end{remark}
Thanks to the commuting relations, every monomial in the generators with outputs in $\{0,1\}$ can first be written in the form
\begin{align*}
q=M_{a_1}^{y_1}\cdots M_{a_r}^{y_r}
N_{b_1}^{x_1}\cdots N_{b_s}^{x_s}.
\end{align*}
The PVM relations reduce any pair of adjacent generators belonging to the same PVM. 
The nonzero reduced words, including $I$, are linearly independent by the free-product normal form. For an operator word $q$ in these generators, let $\operatorname{red}(q)$ denote the result of the PVM and commuting reductions, which is either zero or a reduced word, so that \eqref{eq301} is equivalent to
\begin{align*}
\lambda I-\Phi_{H_2}=\sum_{i,j=0}^{24} z_{ij}\operatorname{red}(w_i^*w_j).
\end{align*}
For each nonzero reduced word $q$, define
\begin{align*}
c_Z(q)=\sum_{\substack{0\leq i,j\leq24\\
\operatorname{red}(w_i^*w_j)=q}}z_{ij}.
\end{align*}
By \eqref{eq30}, the target coefficients are
\begin{align}
\label{eq304}c_Z(I) &= \lambda, \\
\label{eq305}c_Z(M_a^y) &= -\frac14, \qquad y,a\in\{0,1\}, \\
\label{eq306}c_Z(N_b^x) &= -\frac14, \qquad x,b\in\{0,1\},
\end{align}
and, for $x,y,a,b\in\{0,1\}$,
\begin{align}
\label{eq307}c_Z(M_a^yN_b^x) = \begin{cases}
\dfrac14, & (a,b)\in \bigl\{(x,y),(x,1-y),(1-x,y)\bigr\}, \\[2mm]
0, & (a,b)=(1-x,1-y). \end{cases}
\end{align}
Every other reduced word has coefficient zero.

For fixed $\lambda$, these coefficient conditions \eqref{eq304}--\eqref{eq307}, together with the vanishing of all remaining coefficients, form a finite linear system in the entries of $Z$, with the additional requirement that $Z\geq0$. The full system and its derivation are given in Appendix~\ref{app:coefficient-equations}.

\begin{remark}\rm 
    The preceding word reduction shows that the identity
\begin{equation*}
    \lambda I-\Phi_{H_2}=\mathbf w^*(Z\otimes1)\mathbf w
\end{equation*}
holds if and only if all the linear equations \eqref{eq310}--\eqref{eq318} recorded in Appendix~\ref{app:coefficient-equations} hold for all indices satisfying the stated conditions.
\end{remark}

\subsection{Symmetry and block decomposition}\label{sec:symmetry}

Let
\begin{equation*}
\mathcal W=\operatorname{span}_{\mathbb R}\{w_0,w_1,\cdots,w_{24}\}\subseteq\mathcal A.
\end{equation*}
We identify $\mathcal{W}$ with $\mathbb R^{25}$ by letting the standard vector $e_i$ correspond to the operator $w_i$. There exist $*$-automorphisms $\widetilde{\alpha},\widetilde{\beta},\widetilde{\tau}$ on $\mathcal{A}$ satisfying
\begin{align*}
\widetilde\alpha(M_a^y)
&=M_{1-a}^y,
&\widetilde\alpha(N_b^x)
&=N_b^{1-x},
\\
\widetilde\beta(M_a^y)
&=M_a^{1-y},
&\widetilde\beta(N_b^x)
&=N_{1-b}^x,
\\
\widetilde\tau(M_a^y)
&=N_a^y,
&\widetilde\tau(N_b^x)
&=M_b^x.
\end{align*}
Here $x,y,a,b\in\{0,1\}$; the images of the pass projections are determined by the PVM relations. These automorphisms preserve the Bell operator $\Phi_{H_2}$. We denote their restrictions to $\mathcal W$ by $\alpha,\beta,\tau$, respectively. Moreover,
\begin{align*}
\alpha^2=\beta^2=\tau^2&=\operatorname{id},&\alpha\beta&=\beta\alpha,&\tau\alpha\tau&= \beta.
\end{align*}
Hence
\begin{equation*}
    G_{\mathrm{sym}}=\langle\alpha,\beta,\tau\rangle \cong D_8.
\end{equation*}
This provides a natural action of the dihedral group $D_8$ of order eight on $\mathcal W$. Let $\operatorname{Perm}(25)$ denote the group of all $25\times25$ permutation matrices. Every $\sigma\in G_{\mathrm{sym}}$ permutes the words in \eqref{eq:word-vector}, and we define $Q_\sigma\in\operatorname{Perm}(25)$ by $Q_\sigma |j\ra=|\sigma(j)\ra$, where the correspondence $j\mapsto \sigma(j)$ is defined via $\sigma(w_j)=w_{\sigma(j)}$. Then \begin{align*} 
(\operatorname{id}\otimes\sigma)(\mathbf w) = (Q_\sigma^T\otimes\operatorname{id})\mathbf w.  
\end{align*} Moreover, the map $\sigma\longmapsto Q_\sigma$ is a representation of $G_{\mathrm{sym}}\cong D_8$. 
\begin{lemma}\label{lem:averaging}
    If there exists a real symmetric matrix $Z$ satisfying 
\begin{align}
\lambda I-\Phi_{H_2}=\mathbf w^*(Z\otimes1)\mathbf w,
\label{eq:coefficient-identity-before-averaging}
\end{align}
then the averaged matrix 
\begin{align*}
\mathcal R(Z)=\frac18
\sum_{\sigma\in G_{\mathrm{sym}}}
Q_\sigma^T Z Q_\sigma,
\end{align*} also satisfies \eqref{eq:coefficient-identity-before-averaging}. 
\end{lemma}
\begin{proof}
    For each $\sigma\in G_{\mathrm{sym}}$, let $\widetilde{\sigma}$ denote the corresponding $*$-automorphism on $\mathcal A$. A crucial observation is that, for any $\sigma\in G_{\rm sym}$,
        $$\widetilde{\sigma}(\Phi_{H_2})=\Phi_{H_2},$$
    since $\widetilde{\alpha}(\Phi_{H_2}) = \widetilde{\beta}(\Phi_{H_2})=\widetilde{\tau}(\Phi_{H_2})=\Phi_{H_2}$ and $\widetilde{\sigma}$ is generated by $\{\widetilde{\alpha},\widetilde{\beta},\widetilde{\tau}\}$. Now if $Z$ is real symmetric and satisfies \eqref{eq:coefficient-identity-before-averaging}, then 
    \begin{align*}
   {\bf w}^*(Q_{\sigma}^TZQ_{\sigma}\otimes 1){\bf w}&=(\operatorname{Id}\otimes \sigma^{-1})({\bf w})^*(Z\otimes 1)(\operatorname{Id}\otimes \sigma^{-1})({\bf w})\\
    & = \widetilde{\sigma}^{-1}({\bf w}^*(Z\otimes 1){\bf w})
    =\widetilde{\sigma}^{-1}(\lambda I-\Phi_{H_2})=\lambda I-\Phi_{H_2}
\end{align*}
for all $\sigma\in G_{\text{sym}}$. Averaging over $G_{\mathrm{sym}}$ proves the assertion. 
\end{proof}

\begin{remark}
    \rm The averaged matrix $\mathcal R(Z)$ in Lemma \ref{lem:averaging} is real symmetric and satisfies
\begin{align*}
Q_\sigma^T \mathcal R(Z) Q_\sigma=\mathcal R(Z),\qquad \sigma\in G_{\mathrm{sym}}.
\end{align*}
We call this condition {\it $D_8$-invariance}. Averaging also preserves positive semidefiniteness. Hence, in seeking a positive semidefinite solution to \eqref{eq:coefficient-identity-before-averaging}, we may restrict attention to $D_8$-invariant matrices $Z\in M_{25}(\re)_{sym}$.

\end{remark}

\par 
The representation theory of the dihedral group $D_8$ allows a further reduction of the matrix problem. Up to unitary equivalence, there are exactly five irreducible unitary representations
\[
\chi_{+,+},\chi_{+,-},\chi_{-,+},\chi_{-,-},E,
\]
where the first four are one-dimensional and $E$ is two-dimensional; see \cite[Section~5.3, pp.~36--37]{Ser1977}. Their explicit matrices and the associated projection formulas are given in Appendix~\ref{app:representations}.

Every finite-dimensional representation of $D_8$ decomposes as a direct sum of copies of these irreducible representations. Specifically, there exists an invertible matrix $V\in M_{25}(\re)$ and natural numbers $n_{\varepsilon,\delta}$ and $n_E$ such that
\begin{align}\label{eq-decomposition}
V^{-1}Q_{\sigma}V=\left ( \bigoplus_{\varepsilon,\delta=+,-}\chi_{\varepsilon,\delta}(\sigma) \operatorname{Id}_{n_{\varepsilon,\delta}}\right )\oplus ( E(\sigma)\otimes \operatorname{Id}_{n_E})
\end{align}
for all $\sigma\in G_{\text{sym}}$. In this decomposition, the multiplicities are
\[
(n_{+,+},n_{+,-},n_{-,+},n_{-,-},n_E)=(5,2,4,2,6).
\]
This calculation is explained in more detail in Appendix~\ref{app:irreducibles}. We refer the reader to Appendix~\ref{app:representations} for the explicit construction of the matrix $V$ and a proof of its invertibility.
\begin{proposition}\label{prop:block-decomposition}
    Let $Z\in M_{25}(\re)_{\text{sym}}$ be $D_8$-invariant and $V\in M_{25}(\mathbb{R})$ be the invertible matrix from \eqref{eq-decomposition}. Then $V^TZV$ has the block-diagonal form
\begin{equation*}
B_{+,+}\oplus B_{+,-}\oplus B_{-,+}\oplus B_{-,-}\oplus B_E\oplus B_E.    
\end{equation*}
Moreover, $Z\geq0$ if and only if all five blocks are positive semidefinite.
\end{proposition}    
\begin{proof}
    Let $D=V^T ZV$. For each $\sigma\in G_{\mathrm{sym}}$, write $R_\sigma=V^{-1}Q_{\sigma}V$. Then, since $VR_\sigma=Q_{\sigma}V$ and $Q_\sigma^TZQ_\sigma=Z$ for all $\sigma$, we have
\begin{align*}
R_\sigma^TDR_\sigma &= (VR_\sigma)^TZ(VR_\sigma)= (Q_\sigma V)^TZ(Q_\sigma V) \notag\\
&= V^TQ_\sigma^TZQ_\sigma V = V^TZV = D
\end{align*}
for all $\sigma\in G_{\mathrm{sym}}$. In addition, by \eqref{eq-decomposition}, we have
\begin{align*}
  R_{\sigma}=\chi_{+,+}(\sigma)\operatorname{Id}_5 \oplus \chi_{+,-}(\sigma)\operatorname{Id}_2 \oplus \chi_{-,+}(\sigma)\operatorname{Id}_4 \oplus \chi_{-,-}(\sigma)\operatorname{Id}_2 \oplus  \bigl(E(\sigma)\otimes\operatorname{Id}_6\bigr),
\end{align*}
which implies that $R_\sigma$ is orthogonal. Thus, $R_{\sigma}D=DR_{\sigma}$ for all $\sigma\in G_{\text{sym}}$. Schur's lemma therefore implies that $D$ has the block-diagonal form 
\begin{align*}
D = B_{+,+} \oplus B_{+,-} \oplus B_{-,+} \oplus B_{-,-} \oplus \bigl(\operatorname{Id}_2\otimes B_E\bigr),
\end{align*}
where all five blocks $B_{+,+}, B_{+,-}, B_{-,+}, B_{-,-},B_E$ are real symmetric matrices of sizes $5$, $2$, $4$, $2$, $6$, respectively.  Consequently,
\begin{align*}
V^TZV=B_{+,+} \oplus B_{+,-} \oplus B_{-,+} \oplus B_{-,-} \oplus B_E \oplus B_E \in M_{25}(\re)_{\text{sym}}
\end{align*}
and, since $V$ is invertible,
\begin{equation*}
Z\geq0 \quad\Longleftrightarrow\quad
B_{+,+},\, B_{+,-},\, B_{-,+},\, B_{-,-},\, B_E \geq 0.\qedhere
\end{equation*}
\end{proof}

\begin{remark}\rm 
    We note that, with the change of basis determined by $V$, the search over real symmetric $D_8$-invariant matrices reduces to the search over the five real symmetric blocks 
\begin{equation*}
    B_{+,+},B_{+,-},B_{-,+},B_{-,-},B_E.
\end{equation*}
Thus only $15+3+10+3+21=52$ independent real parameters remain, compared with $25\cdot26/2=325$ before imposing $D_8$-invariance.
We refer the reader to Appendix~\ref{app:blocks} for the explicit construction of the blocks $B_{+,+}$, $B_{+,-}$, $B_{-,+}$, $B_{-,-}$, $B_E$.
\end{remark}

\subsection{Proof of the main theorem}\label{sec-upper2}
The construction and verification in Appendix~\ref{app:certificate} give the following certificate.

\begin{proposition}\label{prop:exact-certificate}
Let $\lambda=(15+\sqrt{33})/36$, and let $V$ be the matrix from \eqref{eq-decomposition}. There exist real symmetric positive semidefinite blocks $B_{+,+},B_{+,-},B_{-,+},B_{-,-},B_E$ of sizes $5,2,4,2$, and $6$, respectively, such that the matrix
\begin{align}
Z_\ast = (V^{-1})^T \Bigl( B_{+,+} \oplus B_{+,-} \oplus B_{-,+} \oplus B_{-,-} \oplus B_E \oplus B_E \Bigr) V^{-1}
\label{eq:explicit-definition-Z-star}
\end{align}
is positive semidefinite and satisfies
\begin{align}
\lambda I-\Phi_{H_2} = \mathbf w^*(Z_\ast\otimes1)\mathbf w.
\label{eq:exact-coefficient-identity}
\end{align}
\end{proposition}
\begin{proof}
The blocks constructed in Appendix~\ref{app:blocks} are shown to be positive semidefinite in Appendix~\ref{app:verification}. Since $V$ is invertible, \eqref{eq:explicit-definition-Z-star} gives $Z_\ast\geq0$.

Appendix~\ref{app:verification} also verifies that the entries of $Z_\ast$ satisfy the coefficient equations \eqref{eq310}, \eqref{eq311}--\eqref{eq312}, and \eqref{eq315}--\eqref{eq318} in Appendix~\ref{app:coefficient-equations}. These equations express precisely the conditions \eqref{eq304}--\eqref{eq307} for $Z=Z_\ast$, together with the vanishing of all remaining reduced-word coefficients.

As established in Section~\ref{sec-upper1}, the nonzero reduced words are linearly independent. Hence the preceding coefficient conditions are equivalent to the identity \eqref{eq:exact-coefficient-identity}, which completes the proof.
\end{proof}

\begin{proof}[Proof of the main theorem]
Proposition~\ref{prop:exact-certificate} provides the positive semidefinite matrix $Z_\ast$ satisfying \eqref{eq:certificate-criterion}. The argument in Section~\ref{sec:bell-operator} therefore gives $\omega_{\mathrm{qc}}(H_2)\leq(15+\sqrt{33})/36$. Combining this with Proposition~\ref{prop:quantum-lower-bound} and $\omega_{\rm q}(H_2)\leq\omega_{\mathrm{qc}}(H_2)$, we obtain
\[
\omega_{\rm q}(H_2)=\omega_{\mathrm{qc}}(H_2)=\frac{15+\sqrt{33}}{36}.
\]
\end{proof}

\section{A general upper bound for $\omega_{\mathrm{ns}}(H_n)$}
\label{sec-general-upper-bound}

We conclude by extending the classical upper bound $n/(n+1)$ to all no-signalling strategies and determining the exact
no-signalling value of $H_2$.

Let $X_i=\{0,1\}^{n-1}$ be the input set of player $i\in\{1,\ldots,n\}$, and let $A=\{0,1,P\}$ be the common output set.
A hat configuration $x=(x_1,\ldots,x_n)\in\{0,1\}^n$ is chosen uniformly, and player $i$ receives
\[\pi_i(x)=(x_1,\ldots,x_{i-1},x_{i+1},\ldots,x_n)\in X_i.\]
The team wins if at least one player guesses and every guess is correct.

Recall that a no-signalling strategy over $(X_1,\dots, X_n,A,\dots,A)$ is a conditional probability distribution $p(a\mid z)$, with $a\in A^n$ and $z\in\prod_{i=1}^n X_i$,
whose marginals satisfy
\[\sum_{a_{S^c}\in A^{S^c}}p(a\mid z)=p_S(a_S\mid z_S), \qquad \varnothing\ne S\subseteq\{1,\ldots,n\}, \]
where $a_S=(a_i)_{i\in S}$ and $z_S=(z_i)_{i\in S}$, and the right-hand side is independent of $z_{S^c}$.
We write $p_i=p_{\{i\}}$ for the single-player marginals. The value $\omega_{\mathrm{ns}}(H_n)$ is the supremum of the
average winning probability over all such strategies.

\begin{theorem}\label{thm:general-bound}
For any natural number $n$, we have $\displaystyle \omega_{\mathrm{ns}}(H_n)\leq\frac{n}{n+1}$.
\end{theorem}

\begin{proof}
Fix a no-signalling strategy $p$, and let $s(x)$ be its winning probability at the hat configuration $x\in\{0,1\}^n$.
Every winning output tuple includes at least one correct guess, so the union bound gives
\[s(x)\leq\sum_{i=1}^n p_i(x_i\mid\pi_i(x)).\]

For each $i$, let $x^{(i)}$ be obtained from $x$ by flipping its $i$-th coordinate, i.e., $0\mapsto 1$ and $1\mapsto 0$. Then $\pi_i(x^{(i)})=\pi_i(x)$ for each $i$, and hence the no-signalling conditions imply that player $i$ outputs $x_i$ with the same probability at both configurations $x$ and $x^{(i)}$.
Since this output is \emph{incorrect} at $x^{(i)}$, one  has
\[p_i(x_i\mid\pi_i(x))=p_i(x_i\mid\pi_i(x^{(i)}))\leq 1-s(x^{(i)}).\]
Combining these inequalities yields
\begin{equation}\label{eq400}
s(x)\leq\sum_{i=1}^n\bigl(1-s(x^{(i)})\bigr) = n-\sum_{i=1}^n s(x^{(i)}).
\end{equation}
Summing \eqref{eq400} over $x\in\{0,1\}^n$ and using the fact that $x\mapsto x^{(i)}$ is a bijection for each $i$, one has $\sum_{x} s(x^{(i)}) = \sum_x s(x)$ and hence
\[(n+1)\sum_{x\in\{0,1\}^n}s(x)\leq n2^n.\]
Therefore,
\[\omega(H_n,p)=\frac1{2^n}\sum_{x\in\{0,1\}^n}s(x)\leq\frac{n}{n+1}.\]
Taking the supremum over $p$ and using the inclusions of the strategy classes completes the proof.
\end{proof}

For certain values of $n$, we can determine $\omega_{\mathrm{ns}}(H_n)$ exactly. We begin with the case $n=2$.

\begin{proposition}\label{prop:ns-two-hat}
For the two-player hat game $H_2$, we have $\displaystyle \omega_{\mathrm{ns}}(H_2)=\frac{2}{3}$.
\end{proposition}

\begin{proof}
It follows from Theorem \ref{thm:general-bound} that $\omega_{\mathrm{ns}}(H_2)\leq2/3$. Let $A=\{0,1,P\}$, and let $x,y\in\{0,1\}$ denote Alice's and Bob's hat colors, respectively, so their inputs are $y$ and $x$. For each $(x,y)$, assign probability $1/3$ to each of the output pairs $(x,P)$, $(P,y)$, and $(1-x,1-y)$, and zero to all other output pairs. The conditional probabilities $p(a,b\mid y,x)$ are:
\[
\begin{array}{c|ccccccccc}
(x,y)\backslash(a,b)
 &(0,0)&(0,1)&(0,P)&(1,0)&(1,1)&(1,P)&(P,0)&(P,1)&(P,P)\\
\hline
(0,0)&0&0&\frac13&0&\frac13&0&\frac13&0&0\\[1mm]
(0,1)&0&0&\frac13&\frac13&0&0&0&\frac13&0\\[1mm]
(1,0)&0&\frac13&0&0&0&\frac13&\frac13&0&0\\[1mm]
(1,1)&\frac13&0&0&0&0&\frac13&0&\frac13&0
\end{array}
\]
For every $x,y\in\{0,1\}$, the single-player marginals satisfy
\begin{align*}
p_A(a\mid y)
&:=\sum_{b\in A}p(a,b\mid y,x)=\frac13, \qquad a\in A,\\
p_B(b\mid x)&:=\sum_{a\in A}p(a,b\mid y,x)=\frac13, \qquad b\in A.
\end{align*}
These marginals are independent of the other player's input, so $p$ is no-signalling. Moreover, for every $(x,y)$,
\[p(x,P\mid y,x)+p(P,y\mid y,x)+p(x,y\mid y,x)=\frac13+\frac13+0=\frac23.\]
Thus, we obtain $\omega(H_2,p)=\frac{2}{3}$.
\end{proof}
To summarize, we have inequalities between values of the 2-hat game
    $$\om_{\rm loc}(H_2)=\frac{1}{2}<\om_{\rm q}(H_2)=\om_{\rm qc}(H_2)= \frac{15+\sqrt{33}}{36} < \om_{\rm ns}(H_2)=\frac{2}{3}.$$

For $n=2^m-1$, a well-known classical strategy attains the upper bound $n/(n+1)$. We therefore obtain the following corollary.

\begin{corollary}\label{cor:hamming}
If $n=2^m-1$ with a natural number $m$, then
\[\omega_{\mathrm{loc}}(H_n)=\omega_{\rm q}(H_n)=\omega_{\mathrm{qc}}(H_n)=\omega_{\mathrm{ns}}(H_n)=\frac{n}{n+1}.\]
\end{corollary}

\begin{proof}
For $n=1$, a fixed guess attains $1/2$. For $m\geq2$, a perfect Hamming code yields a classical strategy attaining $n/(n+1)$ \cite{LeSe02}. Theorem \ref{thm:general-bound} and the inclusions of the strategy classes give the claimed equalities.
\end{proof}

Thus no-signalling strategies offer no advantage over classical strategies for $n=3,7,15,\ldots$. For $n=4$, however,
\[\frac{3}{4}=\omega_{\mathrm{loc}}(H_4)\leq \omega_{\mathrm{q}}(H_4)\leq \omega_{\mathrm{qc}}(H_4)\leq\omega_{\mathrm{ns}}(H_4)\leq\frac45.\]
These bounds leave open whether $H_4$ admits a quantum advantage. We plan to investigate this question in future work.

\vspace{10pt}

\emph{Acknowledgements}: G.B. was supported by the National Research Foundation of Korea (NRF) grant funded by the Korea government (MSIT) (No.RS-2025-00561391, No.RS-2024-00413957 and RS-2022-NR069971). S.-G.Y. was supported by the National Research Foundation of Korea (NRF) grant funded by the Korea government (MSIT) (No. RS-2025-00561391, No. RS-2025-10802969 and No. RS-2024-00413957). S.-J. P. is supported by the National Natural Science Foundation of China (grant No.~12401163) and the Department of Science and Technology of Hubei Province (Project No.2025EHA041, Project No.2025AFA044). S.-J. P. also acknowledges the Hongyi Postdoc Fellowship from Wuhan University.

\bibliographystyle{alpha}
\bibliography{QHG2026}
\appendix
\section{Representation-theoretic calculations and the matrix $V$}\label{app:representations}
This appendix contains the representation-theoretic calculations used in Section~\ref{sec:symmetry}, the construction of the invertible matrix $V$ appearing in \eqref{eq-decomposition}, and the proof that it is invertible.

\subsection{Irreducible representations and multiplicities}\label{app:irreducibles}
We use the explicit classification of dihedral representations in \cite[Section~5.3, pp.~36--37]{Ser1977}. Serre denotes the dihedral group of order $2n$ by $D_n$, so our $D_8$ corresponds to $D_4$ in his notation.

Let $\alpha,\beta,\tau$ be the three generators of $G_{\text{sym}}$. Up to unitary equivalence, there are exactly five irreducible unitary representations: 
    \begin{equation*}
    \chi_{+,+},\chi_{+,-},\chi_{-,+},\chi_{-,-},E.
    \end{equation*}
The first four representations $\chi_{\varepsilon,\delta}$ are $1$-dimensional and determined by
    \begin{align*}
        &\chi_{\varepsilon,\delta}(\alpha)=\chi_{\varepsilon,\delta}(\beta)=\varepsilon, \  \text{ and }\ \  \chi_{\varepsilon,\delta}(\tau)=\delta, 
\end{align*}
while $E$ is $2$-dimensional and determined by
\begin{align*}
    &E(\alpha)=\left [\begin{array}{cc}0&1\\ 1&0 \end{array} \right ]&
    &E(\beta)=\left [\begin{array}{cc}0&-1\\ -1&0 \end{array} \right ] &
    &E(\tau)=\left [\begin{array}{cc}1&0\\ 0&-1 \end{array} \right ].
\end{align*} 
Before constructing $V$, we compute the multiplicities $n_{\varepsilon,\delta}$ and $n_E$; see also \cite[Section~2.3, Theorem~4]{Ser1977}. We note that $D_8$ has five conjugacy classes, namely 
\begin{equation*}
\{e\},\quad\{\alpha,\beta\},\quad\{\alpha\beta\},\quad\{\tau,\alpha\beta\tau\},\quad\{\alpha\tau,\beta\tau\},    
\end{equation*}
on which the trace function $\sigma\mapsto \operatorname{Tr}(Q_{\sigma})$ takes the values
\begin{equation*}
25,\qquad 1,\qquad 1,\qquad 5,\qquad 1    
\end{equation*}
respectively. This gives the linear system
\begin{equation}
\begin{bmatrix}
1&1&1&1&2\\
1&1&-1&-1&0\\
1&1&1&1&-2\\
1&-1&1&-1&0\\
1&-1&-1&1&0
\end{bmatrix}
\begin{bmatrix}
n_{+,+}\\
n_{+,-}\\
n_{-,+}\\
n_{-,-}\\
n_E
\end{bmatrix}
=
\begin{bmatrix}
25\\
1\\
1\\
5\\
1
\end{bmatrix},   
\end{equation}
whose solution is $(n_{+,+},n_{+,-},n_{-,+},n_{-,-},n_E)=(5,2,4,2,6)$. We note that this vector is used in Section~\ref{sec:symmetry} for the reduction of the matrix $Z$.
\subsection{Orthogonal projections}\label{app:projections}
We now describe the orthogonal projections onto the representation spaces, using the character projection formula \cite[Section~2.6, Theorem~8]{Ser1977}. For each $\varepsilon,\delta\in\{1,-1\}$, also denoted by $\{+,-\}$, the orthogonal projection $\mathsf P_{\varepsilon,\delta}$ onto the representation space of $\chi_{\varepsilon,\delta}(\sigma)\cdot \operatorname{Id}_{n_{\varepsilon,\delta}}$ and $\mathsf P_{E}$ onto the representation space $E(\sigma)\otimes \operatorname{Id}_{n_E}$ can be explicitly written as
\begin{align*}
&\mathsf P_{\varepsilon,\delta} = \frac18
\sum_{r,s,t\in\{0,1\}}
\varepsilon^{r+s}\delta^t
Q_\alpha^r Q_\beta^s Q_\tau^t 
&\mathsf P_{E}  =\frac1 2 (I-Q_\alpha Q_\beta)
\end{align*}
where $Q_{\alpha},Q_{\beta},Q_{\tau}$ are the associated permutation matrices of the generators $\alpha,\beta,\tau$. We further note that the range of $\mathsf P_E$ is split into the $\pm1$ eigenspaces of $Q_\tau$ by
\begin{align}
\label{eq330}\mathsf P_{E,+}
&=
\frac14
(I-Q_\alpha Q_\beta)(I+Q_\tau),
\\
\label{eq331}\mathsf P_{E,-}
&=
\frac14
(I-Q_\alpha Q_\beta)(I-Q_\tau),
\end{align}
and both projections $\mathsf P_{E,+}$ and $\mathsf P_{E,-}$ have rank six. Furthermore, $\mathsf P_{E,+}$ and $\mathsf P_{E,-}$ satisfy
\begin{align}
\label{eq333}Q_\tau \mathsf P_{E,+}
&=
\mathsf P_{E,+}Q_\tau
=
\mathsf P_{E,+},
\\
Q_\tau \mathsf P_{E,-}
&=
\mathsf P_{E,-}Q_\tau
=
-\mathsf P_{E,-}.
\end{align}
The operators $Q_\alpha$ and $Q_\beta$ interchange the two subspaces:
\begin{align*}
Q_\alpha\mathsf P_{E,+}
&=
\mathsf P_{E,-}Q_\alpha,
&
Q_\alpha\mathsf P_{E,-}
&=
\mathsf P_{E,+}Q_\alpha,
\\
 Q_\beta\mathsf P_{E,+}
&=
\mathsf P_{E,-}Q_\beta,
&
Q_\beta\mathsf P_{E,-}
&=
\mathsf P_{E,+}Q_\beta.
\end{align*}
It also follows from \eqref{eq330} and \eqref{eq331} that
\begin{equation}\label{eq334.5}
Q_\alpha Q_\beta \mathsf P_{E,\pm}=-\mathsf P_{E,\pm}.    
\end{equation}
Furthermore, for any of the orthogonal projections $\mathsf P_{\varepsilon,\delta}$ one has
\begin{align}\label{eq332}
Q_\sigma \mathsf P_{\varepsilon,\delta}&=\mathsf P_{\varepsilon,\delta} Q_\sigma
=\chi_{\varepsilon,\delta}(\sigma)\mathsf P_{\varepsilon,\delta}, \qquad \sigma\in G_{\mathrm{sym}}.
\end{align}

\subsection{Construction of $V$}\label{app:construction-V}
The existence of such a matrix $V$ follows from general representation theory \cite{Ser1977}.
We construct $V$ explicitly below, verify \eqref{eq-decomposition}, and prove its invertibility in Appendix~\ref{app:invertibility}. 
    \par Let $e_0,e_1,\cdots,e_{24}$ be the standard basis vectors of $\mathbb R^{25}$. Choose the following ordered bases for the six
components:
\begin{align*}
V_{+,+}
&=
\left[
\mathsf P_{+,+}e_0,
\mathsf P_{+,+}e_9,
\mathsf P_{+,+}e_{12},
\mathsf P_{+,+}e_1,
\mathsf P_{+,+}e_{10}
\right] \in M_{25,5}(\re),\\
V_{+,-}
&=
\left[
\mathsf P_{+,-}e_{10},
\mathsf P_{+,-}e_1
\right]\in M_{25,2}(\re),\\
V_{-,+}
&=
\left[
\mathsf P_{-,+}e_9,
\mathsf P_{-,+}e_{10},
\mathsf P_{-,+}e_1,
\mathsf P_{-,+}e_{12}
\right]\in M_{25,4}(\re),\\
V_{-,-}
&=
\left[
\mathsf P_{-,-}e_1,
\mathsf P_{-,-}e_{10}
\right]\in M_{25,2}(\re),
\\
V_{E,+}
&=
\left[
\mathsf P_{E,+}e_1,
\mathsf P_{E,+}e_{10},
\mathsf P_{E,+}e_{11},
\mathsf P_{E,+}e_{14},
\mathsf P_{E,+}e_2,
\mathsf P_{E,+}e_9
\right]\in M_{25,6}(\re),\\
V_{E,-}
&=
Q_\alpha V_{E,+}\in M_{25,6}(\re).
\end{align*} 
Furthermore, each $V_{\varepsilon,\delta}$ satisfies
\begin{align}
Q_\sigma V_{\varepsilon,\delta}=\chi_{\varepsilon,\delta}(\sigma)V_{\varepsilon,\delta},
\qquad
\sigma\in G_{\mathrm{sym}},
\label{eq:one-dimensional-block-action}
\end{align}
by \eqref{eq332}, and $V_E=[V_{E,+}~V_{E,-}]\in M_{25,12}(\re)$ satisfies
\begin{align}\label{eq335}
    Q_{\sigma}V_E=V_E(E(\sigma)\otimes \operatorname{Id}_6),\qquad \sigma\in G_{\text{sym}},
    \end{align}
by \eqref{eq333}--\eqref{eq334.5}. Combining \eqref{eq:one-dimensional-block-action} and \eqref{eq335}, we obtain the desired decomposition
\begin{align*}
    Q_{\sigma}V=V\cdot \left (\chi_{+,+}(\sigma)\cdot \operatorname{Id}_{5}\oplus \chi_{+,-}(\sigma)\cdot \operatorname{Id}_{2}\oplus \chi_{-,+}(\sigma)\cdot \operatorname{Id}_{4}\right .
   \left . \oplus \chi_{-,-}(\sigma)\cdot \operatorname{Id}_{2}\oplus ( E(\sigma)\otimes \operatorname{Id}_{6})  \right )
\end{align*}

\medskip

\subsection{Invertibility of $V$}\label{app:invertibility}
We now show that the columns of
\begin{align*}
V=\left[V_{+,+}\ V_{+,-}\ V_{-,+}\ V_{-,-}\ V_{E,+}\ V_{E,-}\right] \in M_{25}(\re)
\end{align*}
are linearly independent. %

\subsubsection{The first four blocks}\label{sss_14} We first consider the four blocks $V_{\varepsilon,\delta}$. Since every
$Q_\sigma$ is a permutation matrix, the group
$G_{\mathrm{sym}}$ acts on the standard basis
$e_0,e_1,\cdots,e_{24}$ of $\mathbb R^{25}$. For each
$i\in\{0,1,\cdots,24\}$, let
\begin{align*}
\mathcal O(i)
=
\left\{
Q_\sigma e_i:
\sigma\in G_{\mathrm{sym}}
\right\}
\end{align*}
be the orbit of $e_i$. With the ordering in \eqref{eq:word-vector}, the action on
$e_0,e_1,\cdots,e_{24}$ has the following five orbits:
$$\mathcal O(0)=\{e_0\},
\quad
\mathcal O(9)= \{e_9,e_{15},e_{18},e_{24}\},
\quad
\mathcal O(12)= \{e_{12},e_{14},e_{19},e_{21}\},$$
$$\mathcal O(1)=\{e_1,e_2,\cdots,e_8\},\quad 
\mathcal O(10)=\{e_{10},e_{11},e_{13},e_{16},
e_{17},e_{20},e_{22},e_{23}\}.$$

We will focus on the five representative vectors $e_0,e_9,e_{12},e_1,e_{10}$. For each $i\in \left\{0,\cdots,24\right\}$, let
\begin{align*}
G_i=\left\{\sigma\in G_{\mathrm{sym}}:Q_\sigma e_i=e_i\right\}
\end{align*}
be the stabilizer of $e_i$. Specifically, we have
\begin{align*}
    G_0=G_{\text{sym}}, \ \ 
    G_9=\left\{e,\tau\right\}, \ \ \ 
    G_{12}=\left\{e,\alpha\beta\tau\right\}, \ \ 
    G_{1}=\left\{e\right\}, \ \ 
    G_{10}=\left\{e\right\}.
\end{align*}
This calculation shows 
\begin{equation*}
    \left\{\begin{array}{cll}
    \chi_{-,+}\equiv 1&\text{ on }G_9,G_{12},G_1,G_{10}\\
    \chi_{-,+}\not\equiv 1& \text{ on }G_0
    \end{array} \right .
\end{equation*}

Now let us compute $\mathsf P_{-,+}e_i$ for the representative vectors $e_i$. For each $i$, let $T_i$ be a set of left coset representatives such that
\begin{equation*}
    G_{\text{sym}}=\bigcup_{\rho\in T_i}\rho G_i=T_iG_i,
\end{equation*}
and $\rho G_i\cap \rho'G_i\neq \emptyset$ implies $\rho=\rho'\in T_i$. Then we have
\begin{align*}
\mathsf P_{-,+}e_i&=\frac18\sum_{r,s,t\in\{0,1\}} (-1)^{r+s} Q_\alpha^rQ_\beta^sQ_\tau^t e_i=\frac18\sum_{\sigma\in G_{\text{sym}}} \chi_{-,+}(\sigma) Q_\sigma e_i\\
&=\frac{1}{8}\sum_{\rho\in T_i}\sum_{h\in G_i}\chi_{-,+}(\rho h)Q_{\rho h}e_i=\frac{1}{8}\sum_{\rho\in T_i}\sum_{h\in G_i}\chi_{-,+}(\rho )\chi_{-,+}(h)Q_{\rho }Q_he_i.
\end{align*}
Since $h\in G_i$ implies $Q_he_i=e_i$, we obtain
\begin{align*}
  \mathsf P_{-,+}e_i=  \frac{1}{8}\left (\sum_{h\in G_i}\chi_{-,+}(h)\right )\sum_{\rho\in T_i}\chi_{-,+}(\rho)Q_{\rho}e_i\in \operatorname{span}\mathcal{O}(i).
\end{align*}
Using the fact
\begin{equation*}
    \sum_{h\in G_i}\chi_{-,+}(h)=\left\{\begin{array}{cll}
    |G_i|&,\text{ if }\chi_{-,+}\equiv 1\text{ on }G_i\\
    0&,\text{ if }\chi_{-,+}\not\equiv 1\text{ on }G_i
    \end{array} \right . ,
\end{equation*}
it follows that if $\chi_{-,+}\equiv 1$ on $G_i$, then
\begin{align*}
\mathsf P_{-,+}e_i=  \frac{1}{|\mathcal{O}(i)|}\sum_{\rho\in T_i}\chi_{-,+}(\rho)Q_{\rho}e_i\in \operatorname{span}\mathcal{O}(i),
\end{align*}
which is a signed average over the orbit $\mathcal O(i)$. Otherwise, we have $\mathsf P_{-,+}e_i=0$. More explicitly, for the five representative vectors $e_i$, we have
\begin{align*}
& \mathsf P_{-,+}e_0=0\\
& \mathsf P_{-,+}e_9
=\frac14\left(e_9-e_{15}-e_{18}+e_{24}\right)\in\operatorname{span}\mathcal{O}(9),
\notag\\
&\mathsf P_{-,+}e_{10}
=
\frac18
\left(
e_{10}-e_{11}+e_{13}-e_{16}
-e_{17}+e_{20}-e_{22}+e_{23}
\right)\in \operatorname{span}\mathcal{O}(10),
\notag\\
&\mathsf P_{-,+}e_1
=
\frac18
\left(
e_1-e_2-e_3+e_4
+e_5-e_6-e_7+e_8
\right)\in \operatorname{span}\mathcal{O}(1),
\notag\\
&\mathsf P_{-,+}e_{12}
=
\frac14
\left(
e_{12}-e_{14}-e_{19}+e_{21}
\right)\in \operatorname{span}\mathcal{O}(12).
\end{align*}
The four nonzero vectors above lie in distinct orbit spans and are therefore mutually orthogonal. Hence
\begin{equation}
V_{-,+}=\left[
\mathsf P_{-,+}e_9,\,
\mathsf P_{-,+}e_{10},\,
\mathsf P_{-,+}e_1,\,
\mathsf P_{-,+}e_{12}
\right]
\end{equation}
has rank $4$. The same argument applies to the other three blocks $V_{\varepsilon,\delta}$. 
\subsubsection{The fifth and sixth blocks}\label{sss_56} It remains to explain the choices of columns in $V_{E,+}$ and $V_{E,-}$. For this purpose, consider the subgroup
\begin{equation*}
H=\langle \alpha\beta,\tau\rangle=\{e,\alpha\beta,\tau,\alpha\beta\tau\} \subseteq G_{\mathrm{sym}}.
\end{equation*}
Since $\alpha\beta$ is central in $G_{\mathrm{sym}}\cong D_8$, the subgroup
$H$ is isomorphic to $\mathbb Z_2\times\mathbb Z_2$. Define a
one-dimensional character $\eta_+$ of $H$ by
\begin{equation*}
\eta_+(\alpha\beta)=-1, \qquad \eta_+(\tau)=1.    
\end{equation*}
Thus,
\begin{equation*}
\eta_+(e)=\eta_+(\tau)=1,
\qquad \eta_+(\alpha\beta) =\eta_+(\alpha\beta\tau) =-1.    
\end{equation*}
The orthogonal projection $\mathsf P_{E,+}$ can therefore be written as
\begin{align*}
\mathsf P_{E,+}
&=\frac14
\sum_{h\in H}\eta_+(h)Q_h=
\frac14 \left( I+Q_\tau-Q_{\alpha\beta}-Q_{\alpha\beta\tau} \right).
\end{align*}

For each $i\in\{0,1,\cdots,24\}$, let $\mathcal O_H(i)=\left\{ Q_h e_i: h\in H \right\}$
be the orbit of $e_i$ under $H$. There are nine orbits: 
$$\mathcal O_H(0)=
\{e_0\},\quad
\mathcal O_H(1)
=
\{e_1,e_4,e_5,e_8\},
\quad 
\mathcal O_H(2)=\{e_2,e_3,e_6,e_7\},$$
$$ \mathcal O_H(9)=
\{e_9,e_{24}\},\quad
\mathcal O_H(10)=\{e_{10},e_{13},e_{20},e_{23}\},
\quad 
\mathcal O_H(11)=
\{e_{11},e_{16},e_{17},e_{22}\},$$
$$
\mathcal O_H(12)=\{e_{12},e_{21}\},
\quad
\mathcal O_H(14)=\{e_{14},e_{19}\},
\quad 
\mathcal O_H(15)=\{e_{15},e_{18}\}.$$

Consider the nine representative vectors $\left\{e_0,e_1,e_2,e_9,e_{10},e_{11},e_{12},e_{14},e_{15}\right\}$ and for each $i\in \left\{0,\cdots,24\right\}$, let $H_i=\left\{h\in H:Q_he_i=e_i\right\}$ be the stabilizer of $e_i$. We further note that the stabilizers are explicitly given by

$$H_0=H,\qquad H_1=H_2=H_{10}=H_{11}=\{e\},$$
$$H_9=H_{14}=\{e,\tau\},\qquad H_{12}=H_{15}=\{e,\alpha\beta\tau\}.$$
This calculation shows 
\begin{equation*}
    \left\{\begin{array}{cll}
    \eta_{+}\equiv 1&\text{ on }H_1,\ H_2,\ H_9,\ H_{10},\ H_{11},\ H_{14}\\
    \eta_{+}\not\equiv 1& \text{ on }H_0,\ H_{12},\ H_{15}
    \end{array}. \right . 
\end{equation*}

If $T_i$ is a set of representatives for the
cosets of $H_i$ in $H$, then the same calculation as in Appendix~\ref{sss_14} gives
\begin{align*}
\mathsf P_{E,+}e_i = \frac14
\sum_{\rho\in T_i}
\sum_{h\in H_i}
\eta_+(\rho h)Q_{\rho h}e_i=
\frac14
\left(
\sum_{h\in H_i}\eta_+(h)
\right)
\sum_{\rho\in T_i}
\eta_+(\rho)Q_\rho e_i,
\end{align*}
and we have 
\begin{equation*}
    \sum_{h\in H_i}\eta_{+}(h)=\left\{\begin{array}{cll}
    |H_i|&,\text{ if }\eta_{+}\equiv 1\text{ on }H_i\\
    0&,\text{ if }\eta_{+}\not\equiv 1\text{ on }H_i
    \end{array} \right . .
\end{equation*}
Consequently, $\mathsf P_{E,+}e_i=0$ if $\eta_+\not\equiv 1$ on $H_i$. If $\eta_+\equiv 1$ on
$H_i$, then
\begin{align*}
\mathsf P_{E,+}e_i
=
\frac{1}{|\mathcal O_H(i)|}
\sum_{\rho\in T_i}
\eta_+(\rho)Q_\rho e_i \in \operatorname{span}\mathcal{O}_H(i).
\end{align*}
Thus, we focus on the six representative vectors $e_1,e_2,e_9,e_{10},e_{11},e_{14}$ whose signed averages are given explicitly by 
\begin{align*}
&\mathsf P_{E,+}e_1
=
\frac14
\left(
e_1-e_4+e_5-e_8
\right)\in \operatorname{span}\mathcal{O}_H(1),
\notag\\
&\mathsf P_{E,+}e_{10}
=
\frac14
\left(
e_{10}+e_{13}-e_{20}-e_{23}
\right)\in \operatorname{span}\mathcal{O}_H(10),
\notag\\
&\mathsf P_{E,+}e_{11}
=
\frac14
\left(
e_{11}-e_{16}+e_{17}-e_{22}
\right)\in \operatorname{span}\mathcal{O}_H(11),
\notag\\
&\mathsf P_{E,+}e_{14}
=
\frac12
\left(
e_{14}-e_{19}
\right)\in \operatorname{span}\mathcal{O}_H(14),
\notag\\
&\mathsf P_{E,+}e_2
=
\frac14
\left(
e_2-e_3+e_6-e_7
\right)\in \operatorname{span}\mathcal{O}_H(2),
\notag\\
&\mathsf P_{E,+}e_9
=
\frac12
\left(
e_9-e_{24}
\right)\in \operatorname{span}\mathcal{O}_H(9).
\end{align*}
Since the selected orbits $\mathcal{O}_H(i)$ are mutually disjoint, the six nonzero vectors above are mutually orthogonal, and hence
\begin{equation*}
V_{E,+}=\left[
\mathsf P_{E,+}e_1,\,
\mathsf P_{E,+}e_{10},\,
\mathsf P_{E,+}e_{11},\,
\mathsf P_{E,+}e_{14},\,
\mathsf P_{E,+}e_2,\,
\mathsf P_{E,+}e_9
\right] 
\end{equation*}
has rank $6$. Moreover, since $Q_{\alpha}$ is orthogonal, 
$V_{E,-}=Q_\alpha V_{E,+}$ also has rank $6$, and the column space of $V_{E,-}$ is orthogonal to the column space of $V_{E,+}$ since they are contained in the $-1$ and $+1$
eigenspaces of $Q_\tau$, respectively. It follows that $V_E=\left[ V_{E,+}\ V_{E,-} \right]$ has rank twelve. 

Finally, note that the six projections
\begin{equation*}
    \mathsf P_{+,+},\quad
\mathsf P_{+,-},\quad
\mathsf P_{-,+},\quad
\mathsf P_{-,-},\quad
\mathsf P_{E,+},\quad
\mathsf P_{E,-}
\end{equation*}
are pairwise orthogonal and sum to the identity. By Appendices~\ref{sss_14} and~\ref{sss_56}, the six blocks have ranks
\begin{equation*}
    5,\qquad 2,\qquad 4,\qquad 2,\qquad 6,\qquad 6,
\end{equation*}
respectively and thus the matrix
\begin{equation*}
V=\left[V_{+,+}\ V_{+,-}\ V_{-,+}\ V_{-,-}\ V_{E,+}\ V_{E,-} \right]    
\end{equation*}
has rank $25$ and is therefore invertible.
\section{Coefficient equations and the exact certificate}\label{app:certificate}
This appendix gives the full coefficient equations from Section~\ref{sec-upper1}, the explicit positive semidefinite blocks, and the verification of the certificate used in Section~\ref{sec-upper2}.

\subsection{Explicit coefficient equations}\label{app:coefficient-equations}

In this subsection, we translate equations \eqref{eq304}--\eqref{eq307}, together with the vanishing of all remaining coefficients, into an explicit system of linear equations for the entries of a $25\times25$ real symmetric matrix $Z=(z_{ij})_{0\leq i,j\leq 24}\in M_{25}(\re)_{sym}$.

For $x,y,a,b\in\{0,1\}$, define the index labels $\iota_A(y,a)$, $\iota_B(x,b)$ and $\iota_{AB}(y,a,x,b)$ by
\begin{align}\label{eq-iotaAB}
w_{\iota_A(y,a)}&= M_a^y, & w_{\iota_B(x,b)} &= N_b^x, & w_{\iota_{AB}(y,a,x,b)} &= M_a^yN_b^x.
\end{align}
Thus $\iota_A$ labels the Alice words, $\iota_B$ the Bob words, and $\iota_{AB}$ the mixed words. Explicitly, the ordering in \eqref{eq:word-vector} gives
\begin{align*}
\iota_A(y,a)&=1+2y+a,\\
\iota_B(x,b)&=5+2x+b,\\
\iota_{AB}(y,a,x,b)&=9+8y+4a+2x+b.
\end{align*}
Let us compute the coefficient map $c_Z$ for all possible types of
reduced word occurring among the products $w_i^*w_j$. 

\medskip
\noindent\textbf{The identity word.}
\par\smallskip

The only product reducing to $I$ is $I^*I$. Therefore,
\begin{align}\label{eq310}
\lambda=c_Z(I) = z_{00}.
\end{align}

\medskip
\noindent\textbf{Words of degree one.}
\par\smallskip

Fix $y,a\in\{0,1\}$ and put $i=\iota_A(y,a)$. The products
reducing to $M_a^y$ are
\begin{equation}
I^*M_a^y, \qquad (M_a^y)^*I, \qquad (M_a^y)^*M_a^y.    
\end{equation}
Since $Z$ is symmetric and $(M_a^y)^2=M_a^y$,
\begin{align}
-\frac{1}{4}=c_Z(M_a^y)=2z_{0,i}+z_{i,i}.
\label{eq311}
\end{align}
Similarly, if $j=\iota_B(x,b)$, then
\begin{align}
-\frac{1}{4}=c_Z(N_b^x)=2z_{0,j}+z_{j,j}.
\label{eq312}
\end{align}

\medskip
\noindent\textbf{Mixed words of degree two.}
\par\smallskip

Fix $x,y,a,b\in\{0,1\}$, and write
\begin{equation}
i=\iota_A(y,a),
\qquad
j=\iota_B(x,b),
\qquad
k=\iota_{AB}(y,a,x,b).    
\end{equation}
The reduced word $M_a^yN_b^x$ arises from
\begin{align}
&I^*(M_a^yN_b^x),
\quad (M_a^yN_b^x)^*I, \quad (M_a^y)^*N_b^x, \quad (N_b^x)^*M_a^y,
\\
&(M_a^y)^*(M_a^yN_b^x), \quad (M_a^yN_b^x)^*M_a^y, \quad (N_b^x)^*(M_a^yN_b^x), \\
&(M_a^yN_b^x)^*N_b^x, \quad (M_a^yN_b^x)^*(M_a^yN_b^x). 
\end{align}
Hence
\begin{align}
c_Z(M_a^yN_b^x)=2z_{0,k}+2z_{i,j}+2z_{i,k}+2z_{j,k}+z_{k,k}.
\label{eq315}
\end{align}
By \eqref{eq307}, this coefficient must satisfy
\begin{align}
c_Z(M_a^yN_b^x)=\begin{cases}
\dfrac14,&(a,b)\in
\bigl\{(x,y),(x,1-y),(1-x,y)\bigr\},
\\[2mm]
0,&(a,b)=(1-x,1-y).
\end{cases}
\label{eq:explicit-target-coefficient-mixed}
\end{align}

\medskip
\noindent\textbf{Words containing either two Alice projections or two Bob projections.}
\par\smallskip

Let $y\neq y'$ and fix $a,a'\in\{0,1\}$. The word $M_a^yM_{a'}^{y'}$ arises only from $(M_a^y)^*M_{a'}^{y'}$. Therefore,
\begin{align}
0=c_Z(M_a^yM_{a'}^{y'}) = z_{\iota_A(y,a),\iota_A(y',a')}.
\label{eq316}
\end{align}
Similarly, if $x\neq x'$, then
\begin{align}
0=c_Z(N_b^xN_{b'}^{x'})=z_{\iota_B(x,b),\iota_B(x',b')}.
\label{eq313}
\end{align}

\medskip
\noindent\textbf{Words containing two Alice projections and one Bob projection.}
\par\smallskip

If $y\neq y'$, then
\begin{align}
0&=c_Z(M_a^yM_{a'}^{y'}N_b^x)
\notag\\
&=z_{\iota_A(y,a),\iota_{AB}(y',a',x,b)}+z_{\iota_A(y',a'),\iota_{AB}(y,a,x,b)}+z_{\iota_{AB}(y,a,x,b),\iota_{AB}(y',a',x,b)}.
\label{eq314}
\end{align}

\medskip
\noindent\textbf{Words containing one Alice projection and two Bob projections.}
\par\smallskip

If $x\neq x'$, then
\begin{align}
0&=
c_Z(M_a^yN_b^xN_{b'}^{x'})
\notag\\
&=z_{\iota_B(x,b),\iota_{AB}(y,a,x',b')}+z_{\iota_B(x',b'),\iota_{AB}(y,a,x,b)}+z_{\iota_{AB}(y,a,x,b),\iota_{AB}(y,a,x',b')}.
\label{eq317}
\end{align}

\medskip
\noindent\textbf{Words of degree four.}
\par\smallskip

If $y\neq y'$ and $x\neq x'$, then
\begin{align}
0=c_Z(M_a^yM_{a'}^{y'}N_b^xN_{b'}^{x'})=
z_{\iota_{AB}(y,a,x,b),
   \iota_{AB}(y',a',x',b')}.
\label{eq318}
\end{align}
The formulas \eqref{eq310}--\eqref{eq318} exhaust all nonzero reduced words arising from $w_i^*w_j$, since each $w_i$ contains at most one Alice projection and at most one Bob projection. 
\subsection{Explicit positive semidefinite blocks}\label{app:blocks}
We use the ordered columns of $V$ specified in Appendix~\ref{app:construction-V}.

Let
\begin{align*}
A_{+,+}
&=
\frac1{432}
\begin{bmatrix}
12(15+\sqrt{33}) & 4(21-\sqrt{33}) & 2(21-5\sqrt{33}) \\
4(21-\sqrt{33}) & 6(15-2\sqrt{33}) & 2(9-2\sqrt{33}) \\
2(21-5\sqrt{33}) & 2(9-2\sqrt{33}) & 8(6-\sqrt{33})
\end{bmatrix},
\\[2mm]
X_{+,+}
&=
\begin{bmatrix}
-\dfrac38&0 \\[1mm]
\dfrac{\sqrt{33}-9}{24}&1 \\[1mm]
-\dfrac{\sqrt{33}+3}{24}&-\dfrac12
\end{bmatrix},
\end{align*}
and define
\begin{align*}
B_{+,+}=\begin{bmatrix}
I_3\\
X_{+,+}^T
\end{bmatrix}
A_{+,+}
\begin{bmatrix}
I_3&X_{+,+}
\end{bmatrix}
\in M_5(\re)_{\text{sym}}.
\end{align*}

For the $(+,-)$ component, let
\begin{align*}
A_{+,-}&=\frac1{432}
\begin{bmatrix}
3(\sqrt{33}-4)
\end{bmatrix},  \qquad
X_{+,-}
= \left[ \frac{\sqrt{33}-3}{6} \right],
\end{align*}
and define
\begin{align*}
B_{+,-} = \begin{bmatrix}
1\\
X_{+,-}^T
\end{bmatrix}
A_{+,-}
\begin{bmatrix}
1&X_{+,-}
\end{bmatrix} \in M_2(\re)_{\text{sym}}.
\end{align*}

For the $(-,+)$ component, let
\begin{align*}
A_{-,+}&= \frac1{432}
\begin{bmatrix}
15(\sqrt{33}-5) &6(\sqrt{33}-5) \\
6(\sqrt{33}-5) &3(\sqrt{33}-3) 
\end{bmatrix}, 
\quad X_{-,+} = \begin{bmatrix}
0&\dfrac{5-\sqrt{33}}{2}\\[1mm]
\dfrac{\sqrt{33}-3}{6}&\sqrt{33}-7
\end{bmatrix},
\end{align*}
and define
\begin{align*}
B_{-,+} = \begin{bmatrix}
I_2\\
X_{-,+}^T
\end{bmatrix}
A_{-,+}
\begin{bmatrix}
I_2&X_{-,+}
\end{bmatrix}\in M_4(\re)_{\text{sym}}.
\end{align*}

For the $(-,-)$ component, let
\begin{align*}
B_{-,-} = A_{-,-} = \frac1{432}
\begin{bmatrix}
25-4\sqrt{33}&-3\\
-3&3(2\sqrt{33}-9)
\end{bmatrix}
\in M_2(\re)_{\text{sym}}.
\end{align*}

Finally, let
\begin{align*}
A_E&=\frac1{432}
\begin{bmatrix}
30&3\sqrt{33}-43&0&\sqrt{33}-17\\
3\sqrt{33}-43&75-7\sqrt{33}&\sqrt{33}-9&6\\
0&\sqrt{33}-9&3(3\sqrt{33}-13)&4(2\sqrt{33}-9)\\
\sqrt{33}-17&6&4(2\sqrt{33}-9)&\sqrt{33}+15
\end{bmatrix},
\\[2mm]
X_E&=\begin{bmatrix}
\dfrac{\sqrt{33}-1}{16}&\dfrac{3(\sqrt{33}-9)}{16}\\[2mm]
\dfrac{\sqrt{33}-9}{24}&\dfrac{\sqrt{33}-1}{8}\\[2mm]
\dfrac{9-\sqrt{33}}{24}&\dfrac{\sqrt{33}+3}{8}\\[2mm]
-\dfrac{\sqrt{33}+15}{48}&-\dfrac{\sqrt{33}+7}{16}
\end{bmatrix},
\end{align*}
and define
\begin{align*}
B_E=\begin{bmatrix}
I_4\\
X_E^T
\end{bmatrix}
A_E
\begin{bmatrix}
I_4&X_E
\end{bmatrix}
\in M_6(\re)_{\text{sym}}.
\end{align*}

\subsection{Verification}\label{app:verification}
\subsubsection{Positivity}
We verify positivity using the displayed factorizations of the blocks. By Sylvester's criterion, the five matrices $A_{+,+},A_{+,-},A_{-,+},A_{-,-},A_E$ are positive definite if all their leading principal minors are strictly positive. For $A_{+,+}$, these are
\begin{align*}
\frac{15+\sqrt{33}}{36}, \qquad
\frac{161-17\sqrt{33}}{7776}, \qquad
\frac{23\sqrt{33}-131}{69984},
\end{align*}
and they are all positive. For $A_{+,-}$, the only leading
principal minor is
$\frac{\sqrt{33}-4}{144}>0.$
For $A_{-,+}$, the leading principal minors are
\begin{align*}
\frac{5(\sqrt{33}-5)}{144}, \qquad \frac1{2592},
\end{align*}
and for $A_{-,-}$, they are
\begin{align*}
\frac{25-4\sqrt{33}}{432},
\qquad
\frac{43\sqrt{33}-246}{31104}.
\end{align*}
Finally, the leading principal minors of $A_E$ are
\begin{align*}
\frac5{72}, \qquad \frac{13+6\sqrt{33}}{23328}, \qquad \frac{565-33\sqrt{33}}{6718464}, \qquad \frac{17537-2959\sqrt{33}}{483729408},
\end{align*}
and they are again positive. Hence $A_{+,+},A_{+,-},A_{-,+},A_{-,-}$ and $A_E$ are positive
definite. The fact that
$B_{+,+}, B_{+,-}, B_{-,+}, B_{-,-}, B_E$
are positive semidefinite follows from their factorizations.

\subsubsection{The coefficient identity}
Let $\lambda=(15+\sqrt{33})/36$ and form $Z_\ast$ from the above blocks by \eqref{eq:explicit-definition-Z-star}.
Write $Z_\ast = (z_{ij}^\ast)_{0\leq i,j\leq24}$ and for any nonzero reduced operator word $q$ define
\begin{align*}
c_\ast(q)=\sum_{\substack{0\leq i,j\leq24\\
\operatorname{red}(w_i^*w_j)=q}}z_{ij}^\ast.
\end{align*}
Substituting \eqref{eq:explicit-definition-Z-star} into the coefficient
equations
\eqref{eq310}--\eqref{eq318}
and simplifying in $\mathbb Q(\sqrt{33})$, we obtain
\begin{align*}
c_\ast(I)&=\lambda,\\
c_\ast(M_a^y)&= -\frac14, \qquad y,a\in\{0,1\},\\
c_\ast(N_b^x)&= -\frac14, \qquad x,b\in\{0,1\},
\end{align*}
and
\begin{align*}
c_\ast(M_a^yN_b^x)= \begin{cases}
\dfrac14,& (a,b)\in \bigl\{(x,y),(x,1-y),(1-x,y)\bigr\},
\\[2mm]
0, & (a,b)=(1-x,1-y),
\end{cases}
\end{align*}
for all $x,y,a,b\in\{0,1\}$. Every other reduced word occurring among the products $w_i^*w_j$ has coefficient zero. Comparing these coefficients with \eqref{eq304}--\eqref{eq307}, we conclude that
\begin{align*}
\lambda I-\Phi_{H_2} = \mathbf w^*(Z_\ast\otimes1)\mathbf w.
\end{align*}

\end{document}